\documentclass[iop]{emulateapj} 

\usepackage{graphicx} 
\usepackage{epstopdf}  

\newcommand{\ia}{\'{\i}} 

\begin{document}

\title{Upflows in the central dark lane of sunspot light bridges}

\author{L. Rouppe van der Voort$^1$}\email{rouppe@astro.uio.no} 
\author{L.R. Bellot Rubio$^2$}
\author{A. Ortiz$^1$}

\affil{$^1$ Institute of Theoretical Astrophysics, University of Oslo, P.O. Box 1029 
Blindern, N-0315 Oslo, Norway}
\affil{$^2$ Instituto de Astrof\ia sica de Andaluc\ia a (CSIC), Apdo.\ 3004, 18080 Granada, Spain}

\begin{abstract}
We use high spatial and spectral resolution observations obtained with the CRisp Imaging
SpectroPolarimeter at the Swedish 1-m Solar Telescope to analyze the velocity profile of granular light bridges in a sunspot. 
We find upflows associated with the central dark lanes of the light bridges.
From bisectors in the \ion{Fe}{1}~630.15~nm line we find that the magnitude of the upflows varies with height with the strongest upflows being deeper in the atmosphere. 
Typical upflow velocities measured from the 70\% bisector are around 500~m~s$^{-1}$ with peaks above 1~km~s$^{-1}$.
The upflows in the central dark lane are surrounded by downflows of weaker magnitude, sometimes concentrated in patches with enhanced velocities reaching up to 1.1~km~s$^{-1}$.
A small spatial offset between the upflows and the continuum dark lane is interpreted as a line-of-sight effect due to the elevated nature of the dark lane and the light bridge above the umbral surroundings.
Our observations show that the central dark lane in granular light bridges is not equivalent to the intergranular lanes of normal photospheric granulation that host convective downflows.
These results support recent MHD simulations of magneto-convection in sunspot atmospheres. 
\end{abstract}


\section{Introduction}
\label{sec:introduction}

Light bridges (LBs), long bright structures crossing the dark umbrae of
sunspots, display a large variety in morphology and fine structure.
Some harbor cells that resemble the granulation cells of the normal photosphere (classified as "granular LBs" following 
\citet{Sobotka:1997fr}), 
some are intrusions of penumbral filaments ("filamentary LBs"), while others appear as the alignment of bright grains ("faint granular LBs") and seem to be associated with the isolated umbral dots (UDs) found in the umbra. 
It is now firmly established that LBs are regions harboring convective flows of relatively weakly magnetized plasma in the strongly magnetized surroundings of the umbra. 
Convective flow patterns and motions were measured by e.g.,
\citet{Rimmele:1997vn, 
rimmele08relation, 
2002A&A...383..275H} 
and
\citet{Berger:2003uq}. 
Measurements of the weaker and more inclined magnetic fields in LBs were reported by e.g., 
\citet{1991ApJ...373..683L, 
1995A&A...302..543R, 
Leka:1997ys} 
and \citet{2007PASJ...59S.577K} 
while
\citet{2006A&A...453.1079J} 
find evidence for field-free regions inside a LB.

At high spatial resolution, many granular LBs appear to have a narrow dark lane following the main axis of the LB  \citep{Berger:2003uq,  
Lites:2004zr, 
rimmele08relation, 
2009A&A...504..575S}, 
with small barb-like extensions to the sides. 
The central dark lane has a typical width of the order of 400~km and 
is elevated by a few hundred km above the dark umbral surroundings. 

The 3D MHD simulations of magnetoconvection in an umbral-like atmosphere by 
\citet{2006ApJ...641L..73S} 
show the development of narrow upflow plumes with adjacent downflows. 
The plumes form surface features whose properties agree well with those observed in umbral dots.
Most of the simulated umbral dots have a central dark lane 
that is reminiscent of the dark lanes observed in LBs. The simulations show that the dark lane
is the result of plasma piling up due to strong braking of the vertical flows near the surface. 
The strong surrounding magnetic fields forces the plasma in a
cusp-like region with enhanced density and therefore larger opacity.
This causes the $\tau=1$ surface to be elevated and cut through a
higher, cooler part of the atmosphere resulting in the dark lane in
intensity images.
Nordlund found a similar cusp-shaped central dark lane in a LB formed
in 3D MHD simulations of a field-free gap surrounded by an umbral-like
atmosphere
\citep[cf.][]{2010mcia.conf..243N} 
These simulations predict that, contrary to intuition, the dark lane
in sunspot LBs harbors upflows and is not equivalent to intergranular
lanes of normal granulation that host the downflows of solar surface
convection.
The observational test of this prediction is extremely challenging since it requires high resolution in both the spatial and spectral domains. 
In this Letter we present observations that meet these requirements.

\begin{figure*}
 \includegraphics[width=\columnwidth]{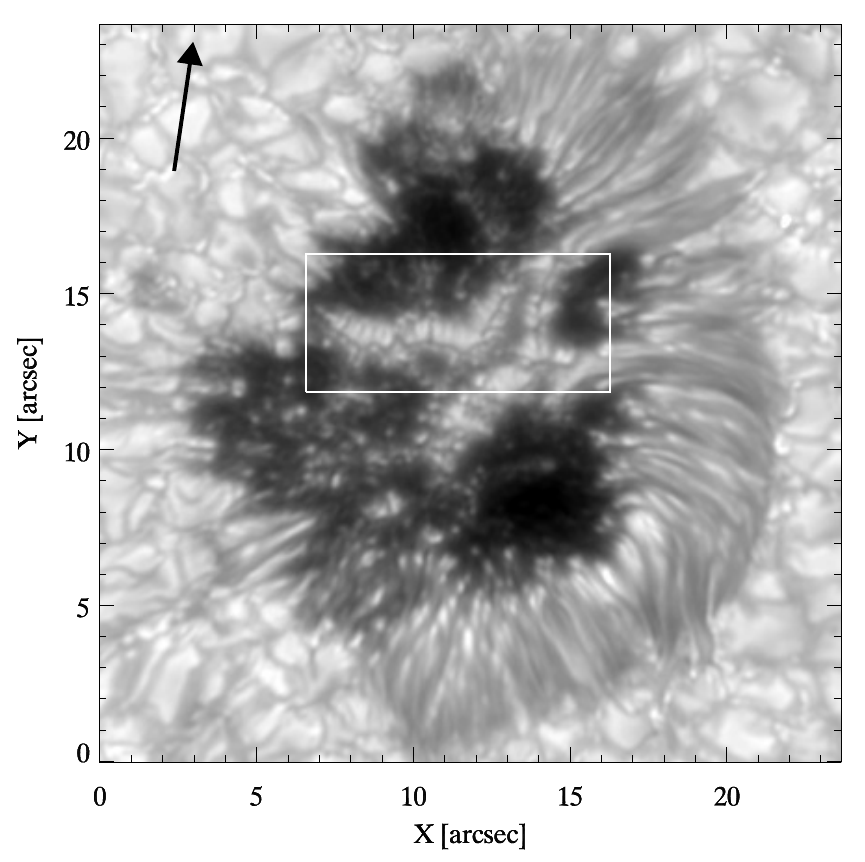}
 \includegraphics[width=\columnwidth]{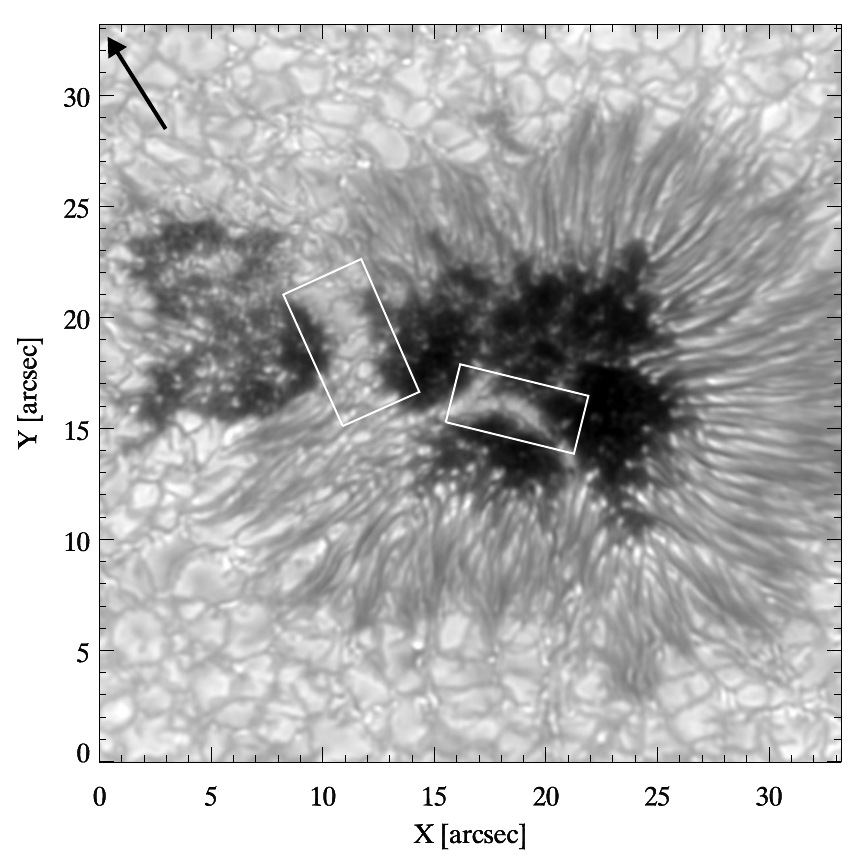}
  \caption{The leading sunspot of AR 11024 observed with CRISP in the continuum at 630.32~nm, on 04-Jul-2009 (left panel) and on 05-Jul-2009 (right panel). The arrows are pointing to disk center. White boxes outline the regions shown in closer detail in Fig.~\ref{fig:lb04Jul} and \ref{fig:lb2_05Jul}.}
    \label{fig:spot}
\end{figure*}

\section{Observations and Data processing}
\label{sec:obs}

The observations were obtained with the CRisp Imaging
SpectroPolarimeter 
\citep[CRISP,][]{2008ApJ...689L..69S} 
installed at the Swedish 1-m Solar Telescope
\citep[SST,][]{2003SPIE.4853..341S} 
on La Palma (Spain).
CRISP is a Fabry-P{\'e}rot interferometer that is designed for diffraction-limited observations in 
the visible ($\lambda/D=0\farcs13$ at 630~nm).
Here we analyze spectro-polarimetric scans of the \ion{Fe}{1}~630.15 and 630.25~nm lines, each sampled at 15 wavelength positions in steps of 4.8 pm, from $-33.6$ to $+33.6$ pm.
At this wavelength, the CRISP transmission profile has a FWHM of 6.4~pm. 
Liquid crystals modulated the light cycling through 4 polarization states and we recorded 9 exposures per state, resulting in a total of 36 exposures per line position. 
A full spectral scan of the two Fe lines and a continuum position at 630.32~nm was completed in 34~s. 

High spatial resolution was achieved by using adaptive optics 
\citep{2003SPIE.4853..370S} 
in combination with the Multi-Object Multi-Frame Blind Deconvolution
\citep[MOMFBD,][]{2005SoPh..228..191V} 
image processing technique.
For the MOMFBD processing, we use simultaneously recorded wide band images as a so-called anchor channel to ensure precise alignment between the sequentially recorded CRISP narrow band images. 
Therefore, the Doppler maps derived from the restored data are virtually free from false signals due to seeing effects.
For further details on the image processing and derivation of Stokes maps, we refer to 
\citet{2008A&A...489..429V}. 

In this Letter we analyze observations from the leading spot of AR 11024 on 04-Jul-2009 and 05-Jul-2009.
Large scale magnetic flux emerged in this region during the observations of 04-Jul-2009 and the spot evolved into a fully developed sunspot with penumbra
\citep[cf.][]{2010A&A...512L...1S} 
  -- one of the few large sunspots of 2009. 
The seeing conditions were very favorable during these two days and from the large volume of recorded data we selected two spectral scans of excellent quality: 
one scan from 04-Jul-2009 recorded at 12:17 UT (heliocentric coordinates $x=-70\arcsec, y=-480\arcsec$, cosine of the observing angle $\mu=0.86$) and one scan from 05-Jul-2009 recorded at 15:05 UT ($x=176\arcsec, y=-480\arcsec$, $\mu=0.84$). 
The spatial resolution in these data sets is close to the diffraction limit of the telescope. 

To obtain a measure of the Doppler velocity as a function of height,
we determine bisectors from the observed Stokes I profiles of
\ion{Fe}{1}~630.15~nm at various intensity levels
(with 0\% representing the line core and 100\% the continuum), as well as the line center position from a parabolic fit to the 5 deepest wavelength points. 
We used the darkest regions in the sunspot umbra as the wavelength reference assuming that the atmosphere in these regions is at rest. 
 For more details, see 
 \citet{2010ApJ...713.1282O}. 

\begin{figure*}
  \includegraphics[width=\columnwidth]{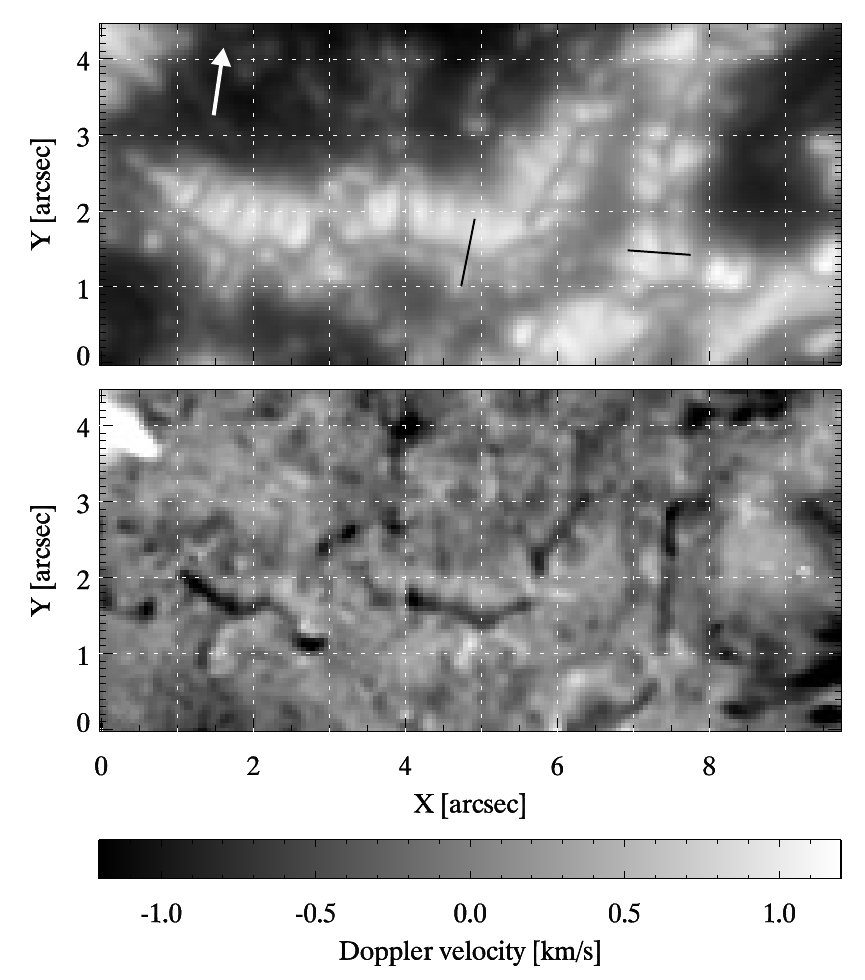}
  \includegraphics[width=\columnwidth]{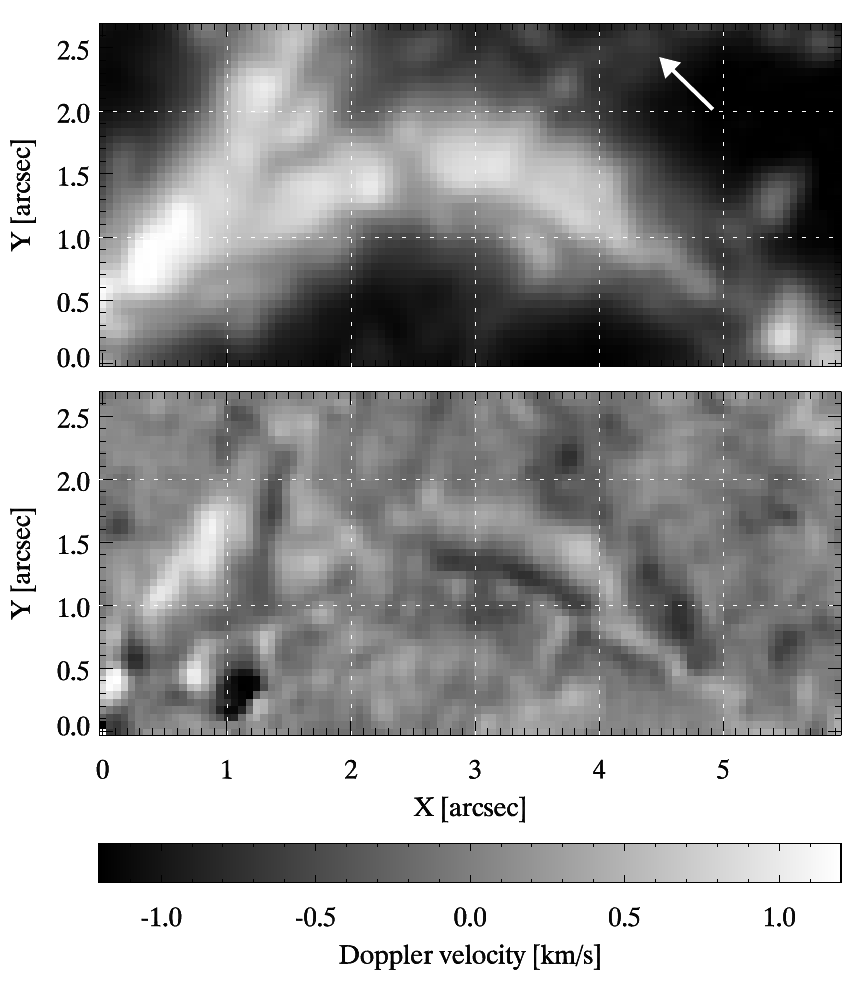}
  \caption{Close-up of the LBs in the sunspot on 04-Jul-2009 and 05-Jul-2009. The top panels are Stokes I maps in the continuum at 630.32~nm. The bottom panels show corresponding maps of the bisector velocity at 70\% of the continuum level. The arrows are pointing to disk center. The thin black lines mark the position of the cross plots shown in Fig.~\ref{fig:cross}. Upflows are dark, the color scaling saturates at $\pm$~1.2~km~s$^{-1}$. }
    \label{fig:lb04Jul}
\end{figure*}

\begin{figure}
  \includegraphics[width=\columnwidth]{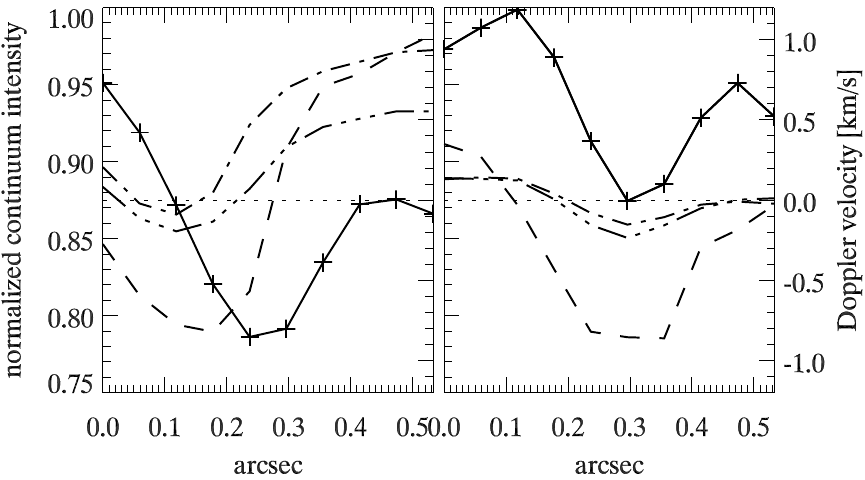}
  \caption{ Cuts across the central dark lane of the LB on 04-Jul-2009
    at the positions marked with thin black lines in the upper left
    panel of Fig.~\ref{fig:lb04Jul}. The $x$-axis of the plot follows
    the up-down direction of the lines. The right panel corresponds to
    the line crossing the LB starting at x=7\arcsec.  The solid line
    with crosses is intensity from the continuum map (left
    $y$-axis). The other three lines are Doppler velocities (right
    $y$-axis): bisectors at 80\% (long dash) and 50\% (dash-triple
    dot), and a parabolic fit to the line core (dash-dot). Negative
    velocities correspond to upflows, the zero point is marked by the
    thin horizontal dotted line.}
    \label{fig:cross}
\end{figure}

\begin{figure}
  \includegraphics[width=\columnwidth]{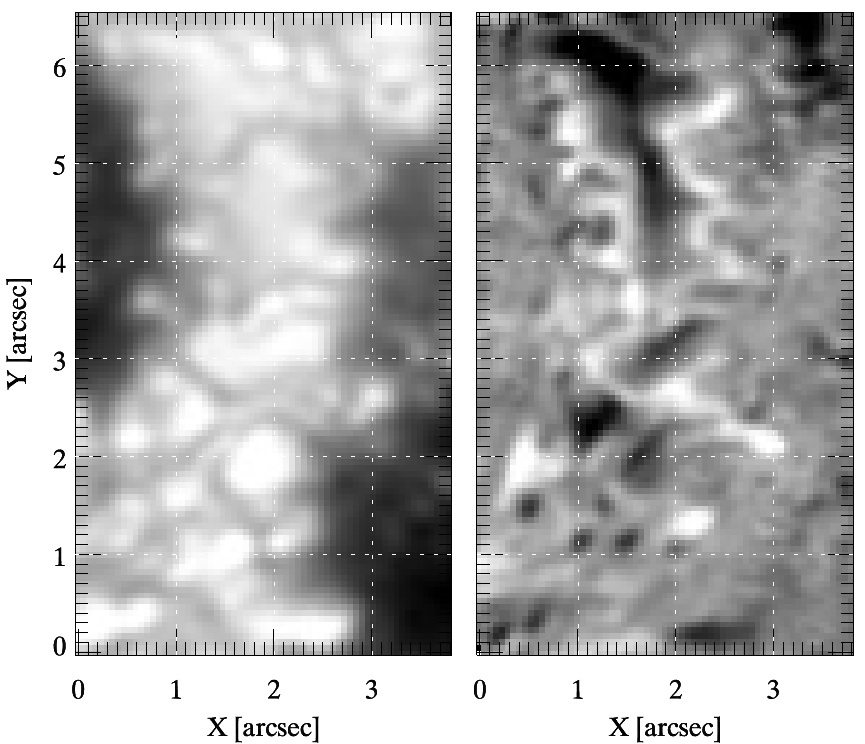}
  \caption{Close-up of the wide LB observed on 05-Jul-2009. The left panel is a Stokes I image in the continuum at 630.32~nm. The right panel shows a map of the bisector velocity at 70\% of the continuum level. Upflows are dark, the color scaling saturates at $\pm$~1.2~km~s$^{-1}$ and has the same linear scaling as for the bisector maps in Fig~\ref{fig:lb04Jul}. }
    \label{fig:lb2_05Jul}
\end{figure}

\section{Results}
\label{sec:results}

Figure~\ref{fig:spot} shows continuum images of the sunspot on both
observing days. On 04-Jul, when the spot was still rapidly growing,
the umbra was dominated by a granular LB that has several branches.
On 05-Jul, the sunspot had grown considerably and two LBs could be
distinguished. One of the LBs separates an umbral part without
penumbra from the rest of the spot.  This LB is considerably wider
than the other LB in the main umbra and does not have a central dark
lane.

In the left part of Fig.~\ref{fig:lb04Jul}, the LB on 04-Jul is shown in closer detail. A central dark lane is clearly visible along most of the LB. 
The corresponding bisector map at 70\% of the continuum shows that for significant parts of the LB, the central dark lane coincides with a narrow upflow region.
The upflow velocities are typically around 500~m~s$^{-1}$ with peaks above 1~km~s$^{-1}$.
The upflow lane in the bisector map is not everywhere exactly centered on the corresponding lane in the continuum but often slightly offset. 
The spatial offset of the Doppler signal appears to be related to the viewing angle: the offset is in the direction towards disk center. 
This is most clearly seen in the part of the LB that is parallel to the $x$-axis, between $x=1\arcsec$ and 6\arcsec.
This is also a region where the size of the bright grains is clearly
asymmetric: the grains on the disk center side are larger than grains
on the opposite side of the dark lane.
The vertically aligned LB at $x=7\farcs5$ is almost parallel to the viewing direction and there we see that the grains have similar sizes on both sides of the dark lane and no spatial offset between the lane in the continuum and Doppler maps. 

The dark lane does not have upflow velocities everywhere -- for
example, between $x=2\farcs5$ to 3\farcs5 it cannot be distinguished
from the background signal in the bisector map.
Sometimes we see narrow upflows where there is no clear dark lane.
The LB shown in the right panel of Fig.~\ref{fig:lb04Jul} illustrates
this: for the upflow lane that is running from $x\approx1\farcs4$,
$y=1\arcsec$ to 2\farcs3 (vertical in the image), no dark lane can be
distinguished in the continuum image.

For most of the dark lane roughly parallel to the $x$-axis 
of the LB in the right of Fig.~\ref{fig:lb04Jul},
corresponding upflows in the bisector map can be found.
In the part of the LB around $x=3\arcsec$ there is a similar spatial
offset of the upflow lane towards disk center with respect to the
position of the dark lane in the continuum.
Here we also see that the sizes of the grains are clearly asymmetric: on the limb side of the dark lane, the grains are much smaller than on the center side. 

When we compare bisector maps at different line depths, we find
stronger upflow signals in the dark lanes closer to the continuum,
i.e., the upflows are more vigorous deeper in the atmosphere.
This is illustrated by the cross plots shown in Fig.~\ref{fig:cross}: in the dark lane, the bisector at 80\% is well below the bisector at 50\% and the line core Doppler velocity. 
The strongest upflow at 80\% in the right panel is more than 800~m~s$^{-1}$, while the strongest upflow at 50\% is only 200~m~s$^{-1}$. 
The slight offset of the strongest upflow with respect to the darkest
point in the continuum dark lane is clearly visible in the left panel.
The offset is about 0\farcs12 
and in the direction towards disk center.  There appears to be a trend
for the Doppler signal from the higher atmospheric layers being
further away from the center of the dark lane in the continuum.
In the right panel, the strongest Doppler signals occur at the
position of the dark lane. This part of the LB is nearly parallel to
the disk center direction (Fig.~\ref{fig:lb04Jul}).

In the bisector maps, the upflow lanes stand out clearly against the background. Close inspection reveals that the lanes are surrounded by downflow regions that coincide with the bright grains of the LB. 
The magnitude of the downflows is smaller than the upflows with typical values around 300~m~s$^{-1}$ and spread out over a larger region as compared to the narrow upflow lanes. 
Small patches of enhanced downflows can be found in the LBs of both days, here the velocities typically go up to $+700$~m~s$^{-1}$.
The plot in the left panel of Fig.~\ref{fig:cross} crosses right
through one of the stronger downflow patches where the velocity
reaches almost $+1$~km~s$^{-1}$.
The strongest downflow is found on 05-Jul at $x=0\farcs8$, $y=1\farcs5$: $+1.1$~km~s$^{-1}$.

Figure~\ref{fig:lb2_05Jul} shows continuum and bisector maps of the wide LB that separates the main umbra from the "naked" umbra without penumbra. 
Unlike the other two LB examples, this LB has no central dark lane and
resembles a strip of normal granulation.
Specially in the upper part of the panels, we see upflow signals in the center where we find the highest intensities and strong downflows at the edges -- just like in normal granulation.

\section{Conclusions}
\label{sec:conclusions}

We find long elongated lanes with 
strong 
upflows associated with the central dark lanes of granular LBs in a large sunspot. 
\citet{rimmele08relation} 
detected upflows in the dark core of a LB with penumbral filament-like appearance, a type of LB that is morphologically quite different from the granular LBs discussed here. 
%
\citet{2008A&A...489..747G} 
report weak upflows of 70~m~s$^{-1}$ in the central dark lane of a LB crossing a small pore.

We find that the magnitude of the upflows varies with height: the Doppler shift of the line core is smaller than the bisector shift near the continuum.
Thus the strongest upflows are measured deeper in the atmosphere, close to the continuum forming region. 
The velocity profile we obtain for the granular LBs, with upflow in the center and downflow towards the edges of the grains, agrees with the idea of the convective nature of LBs.
\citet{rimmele08relation} 
found a similar "bipolar" velocity structure for a filamentary LB and he described this as a flow along the top of an $\Omega$-loop.
Interestingly, for the wide LB shown in Fig.~\ref{fig:lb2_05Jul}, the velocity profile is the same as for the narrower LBs but here we do not observe a central dark lane.
Apparently there is a certain size for which the dark lane ceases to exist.

We interpret the spatial offset of the upflow lanes in the Doppler maps with respect to the dark lane in the continuum as a line-of-sight effect.
Previous observations
\citep{Lites:2004zr, 
rimmele08relation} 
have shown that LBs are elevated above the surrounding umbra. 
We find indeed that the grains are larger on the disk center side of the LB, which indicates that we view the LB from the side and that the limb-ward grains are partly invisible. 
For those parts of the LBs that are parallel to the viewing direction there is no offset between the lanes in the Doppler and continuum maps. 
Only for LB parts with the main axis perpendicular to the viewing
angle we find an offset between the location of maximum LOS velocity
and the center of the dark lane in the continuum.  This offset could
be attributed to the effect of overturning motion in convective flows.
With the upflow plume rising in the center of the LB and downflows
located at the sides, the flow must be more horizontal in between.  If
the LB is observed from the side, like in our observations, the
maximum Doppler shift will not necessarily be observed in the center
of the upflow plume but could well be observed at the side where the
LOS component is larger. In this scenario, the location of maximum
Doppler shift will be at the disk center side of the LB.

Our finding of upflows in the dark lanes of granular LBs supports recent MHD simulations of magneto-convection in umbral atmospheres
\citep{2006ApJ...641L..73S, 
2010mcia.conf..243N}. 
In these simulations, the central dark lane of convective structures embedded in strongly magnetized atmospheres (umbral dots and a field-free strip of granulation) is a cusp-shaped, elevated region with enhanced density as a result of upflowing plasma that is strongly braked at the surface.
Further support for these simulations was recently provided by 
\citet{2010ApJ...713.1282O} 
who measured velocity profiles in umbral dots that agree with the simulations.
Most umbral dots were found to be associated with strong upflows in the deep photospheric layers and concentrated patches of downflows were found at their edges. 

Many observers have pointed at the relation between umbral dots and
LBs 
\citep[see e.g.,][]{Sobotka:1997fr}. 
This is most vividly illustrated by the continuous high-resolution
Hinode observations of the formation of a LB out of umbral dots
emerging from the penumbra and rapidly intruding into the umbra
\citep{2007PASJ...59S.577K}. 
Here we have demonstrated that LBs show upflows in the central dark
lane and downflows surrounding them, much in the same way as umbral
dots \citep{2010ApJ...713.1282O}. 
This supports the idea that both structures are different
manifestations of convection in sunspots, with the downflows
representing the downward component of overturning motions. 
Recent numerical models
 \citep{2006A&A...447..343S, 
2007ApJ...669.1390H, 
2009ApJ...691..640R, 
2009Sci...325..171R} 
indicate that also the filaments of the penumbra are
caused by overturning convection. 
Their dark cores
\citep{2002Natur.420..151S} 
would be the penumbral equivalent of the central dark lane in LBs and
umbral dots, as suggested by the observations of
\citet{2007ASPC..369...71S} 
and
\citet{rimmele08relation}. 
However, this idea needs further scrutiny
because the lateral convective downflows predicted to occur in 
the filaments have not been detected unambiguously as yet.

\acknowledgments
L.B.R. acknowledges financial support from the Spanish Ministerio de Ciencia 
e Innovaci\'on through project AYA2009-14105-C06-06 and from Junta de 
Andaluc\'{\i}a through project P07-TEP-2687. 
A.O. is supported by the Research Council of Norway through grant 177336/V30.
The Swedish 1-m Solar Telescope is operated by the Institute for Solar
Physics of the Royal Swedish Academy of Sciences in the Spanish
Observatorio del Roque de los Muchachos of the Instituto de
Astrof\'{\i}sica de Canarias. 
This research has made use of NASA's Astrophysical Data System.


\end{document}